# Graphene Magnetoresistance Control by Photoferroelectric Substrate

K. Maity, J.-F. Dayen, B. Doudin, R. Gumeniuk, and B. Kundys*

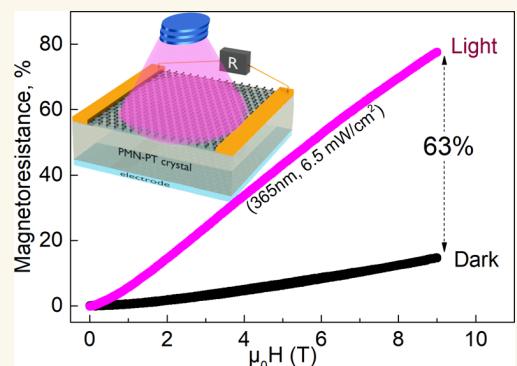

**ABSTRACT:** Ultralow dimensionality of 2D layers magnifies their sensitivity to adjacent charges enabling even postprocessing electric control of multifunctional structures. However, functionalizing 2D layers remains an important challenge for on-demand device−property exploitation. Here we report that an electrical and even fully optical way to control and write modifications to the magnetoresistive response of CVD-deposited graphene is achievable through the electrostatics of the photoferroelectric substrate. For electrical control, the ferroelectric polarization switch modifies graphene magnetoresistance by 67% due to a Fermi level shift with related modification in charge mobility. A similar function is also attained entirely by bandgap light due to the substrate photovoltaic effect. Moreover, an all-optical way to imprint and recover graphene magnetoresistance by light is reported as well as magnetic control of graphene transconductance. These findings extend photoferroelectric control in 2D structures to magnetic dimensions and advance wireless operation for sensors and field-effect transistors.

## 1. INTRODUCTION

The wide range of magnetoresistive (MR) effects has extensive applications across multiple fields and provides valuable insights into the inherent electronic properties of solids. In the case of mono-layer-atomic structures like graphene, the magnetoresistance is connected with nontrivial low-energy electronic excitations of quasi-particle behavior of the charge carriers[1,2] with a record MR value recently reported at room temperature.[3] Because of the high sensitivity of 2D layers to the presence of nearby electric charges, their magnetoresistance can be modified by electric-charge-tunable environments, providing on-demand magnetoresistance control in future 2D-based heterostructures. Indeed, graphene magnetoresistance can be modified by photochemical doping,[4] impurities,[5] or charge polarization.[6] Here we extend and explore the last effect by using a ferroelectric (FE) substrate offering a large doping level modification[7,8] yet with nonvolatile option to imprint multiple memory states with both bulk[8] and surface[9] configurations. Although FE control of graphene resistance is well-known,[10,11] the FE doping effect on magnetoresistive properties remains essentially uncharted. This research direction, however, can offer not only electrical but also optical control possibility. The existence of a photoferroelectric property in FEs makes them outstanding components for 2D hybrid structures, where large changes in charge density can be induced optically.[12−15] Furthermore, a FE can also possess a photovoltaic effect,[16−18] additionally enriching optical functionality even with an all-optical rewritable option.[19] Therefore, single-atom-layer structures with ferroelectric substrates or layers should also manifest magnetoresistance sensitivity to both electrical and optical stimuli in a single device.

## 2. RESULTS AND DISCUSSION

**2.1. Initial Sample Properties.** A comparison of Raman spectra for the PMN-PT30% and Si/SiO$_2$(300 nm) substrates of the same batch is shown in Figure 1.

For the PMN-PT30% substrate the general downshift of the spectrum of ∼16 cm$^{-1}$ is observed due to substrate doping. Indeed, graphene is known to be p-type doped after contact with the PMN-PT.[20] The overall CVD graphene preparation procedure was found to reduce the remanent FE polarization along the [001] direction (Figure 2a) and induce a FE bias shift toward positive electric fields. These FE property modifications can be connected to the thermal history of the FE samples.[21] To verify this, the sample was cooled to 250 K and smaller hysteresis was found at room temperature after this procedure. The polarization ground state of the structure was first characterized electrically to verify ferroelectricity before magnetoresistive measurements. The FE polarization of the substrate (Figure 2b) leads to charge-driven modification in the graphene electrode, forming also the two remanent states of resistance (Figure 2c). While after positive ferroelectric poling (point 1) the graphene resistance manifests increased values (~8.0 kΩ), the negative poling (point 2) testifies to a larger carrier (hole) doping level leading to lower resistance values (~3.0 kΩ). Therefore, in agreement with ref 20, we are in the p-type doping regime, in which a negative electric field application dopes more holes to the graphene, leading to a decrease in resistance for the corresponding remanent ferroelectric state (point 2). Notably, a rather clean graphene layer was obtained after CVD deposition, as confirmed by the Raman spectrum (Figure 1), as well as by the anticlockwise behavior opposite to FE/graphene structures with an additional interfacial dipole layer.[22−30] The clear resistive hysteresis dominated by the ferroelectric polarization is observed, testifying to a strong doping regime far beyond the Dirac point. The two opposite ferroelectric polarization states of the substrate lead to the formation of two remanent states in graphene resistance with nearly 134% change. This change was found to be thermal history dependent, as the FE cycle itself is known to depend on thermal treatment in ferroelectrics.[21] The direct hysteretic behavior with an electric field is observed, indicating no significant interface contribution between graphene and FE: i.e., no hysteresis inversion often reported for other FE-graphene structures.[27−30] The sweeping gate voltage from point 2 to point 1 polarizes the ferroelectric in an upward direction, adding electrons to the graphene channel, which shifts the Fermi level closer to the Dirac point, increasing the resistance of graphene (Figure 2c (inset)). We then reach the saturation of ferroelectric polarization and hence the polarized remanent state (point 1) by returning to zero electric field. For further sweeping of gate voltage on the negative side, we change the ferroelectric polarization to downward, dope graphene with holes, and shift the Fermi level away from the Dirac point. Therefore, at point 2 (remanent state of downward polarization) we get the lower resistance of graphene compared to point 1. So, the ferroelectric polarization variation successfully tunes the wide range of graphene resistance. The nonvolatile behavior of ferroelectric polarization can result in multiple remanent polarization states[31,32] by electrical pulsing of gate voltage: e.g., the graphene resistance can be tuned in the wide range between P1 and P2.[9]

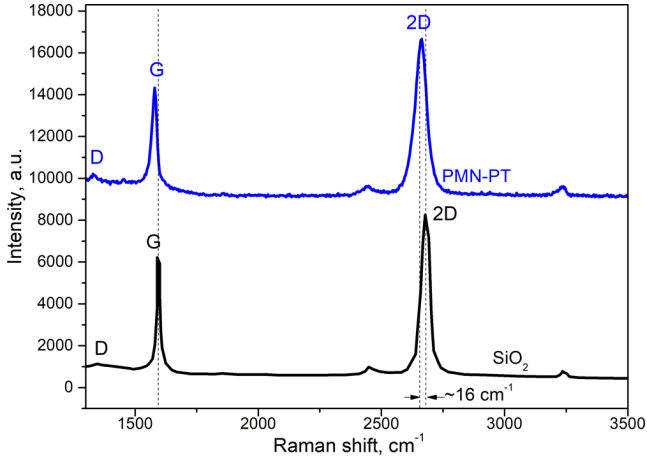

Figure 1. Raman spectra of CVD graphene. The Raman spectrum was measured on a PMN-PT crystal and on the Si/SiO$_2$(300 nm) substrate of the same batch for comparison.

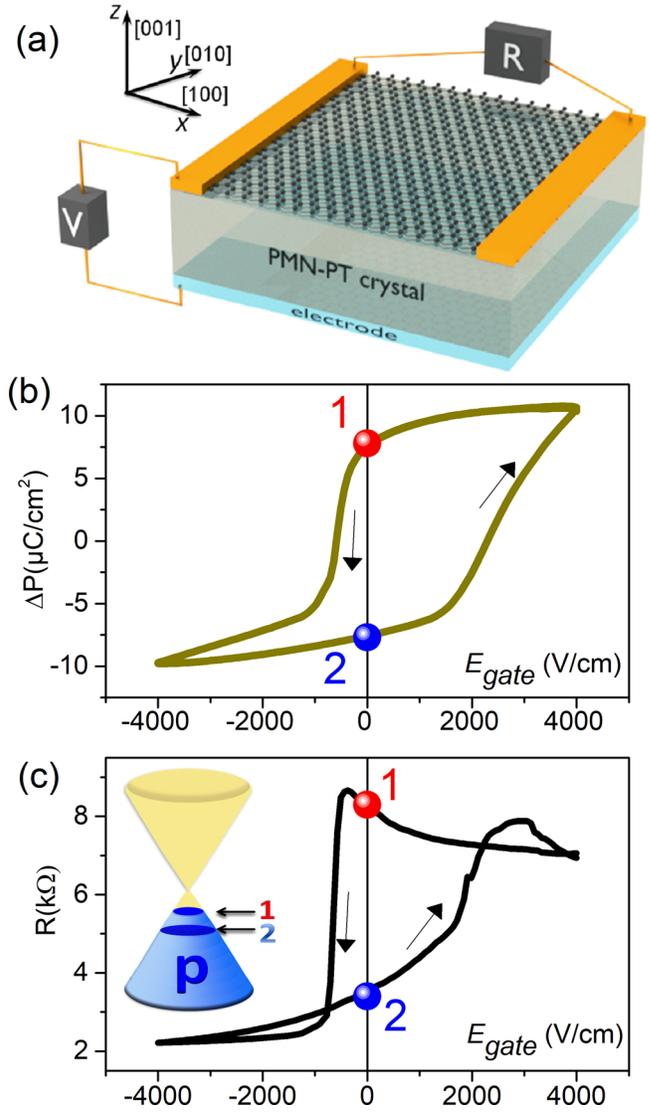

Figure 2. Schematics of the experiment and initial characterization. Geometry of experiment (a). Ferroelectric hysteresis loop along the z-axis (b) and ferroelectrically induced graphene resistive loop measured in situ (c).

The carrier charge density related to the two remanent FE states can be evaluated using the formula $n = 1/Re\mu$ (R is the graphene sheet resistance, e is an electron charge, and μ is the mobility of the charge carrier). The Fermi energy shift can be estimated to be 0.46 eV between the two remanent states (P1 and P2) of a ferroelectric using the relation $E_F = \hbar v_F \sqrt{\pi n}$, with the Fermi velocity of charge carrier $v_F = 1.1 \times 10^6$ m/s.

**2.2. Ferroelectric Control of Magnetoresistance.** To demonstrate FE control of magnetoresistance (MR) defined as $[(R(H) − R(0))/R(0)] \times 100\%$, the curves were recorded at the remanent FE and resistive states 1 and 2 defined by the FE cycle in Figure 2c. Because we operate at room temperature, the magnetoresistance effect in the graphene monolayer can be explained by a classical model, where Lorentz force deviating the path of moving charge carriers leads to resistance increase under applied magnetic field. The MR value is significantly higher at P1 and reaches 80% compared to P2 (~17%) at 300 K at the magnetic field of 9 T. Due to the FE property, the possible minor FE loop existence in the subcoercive region can lead to multiple electrically written remanent states. Therefore, the magnetoresistance property can be predetermined in this way by deterministic electric field poling as schematically shown in Figure 3a.

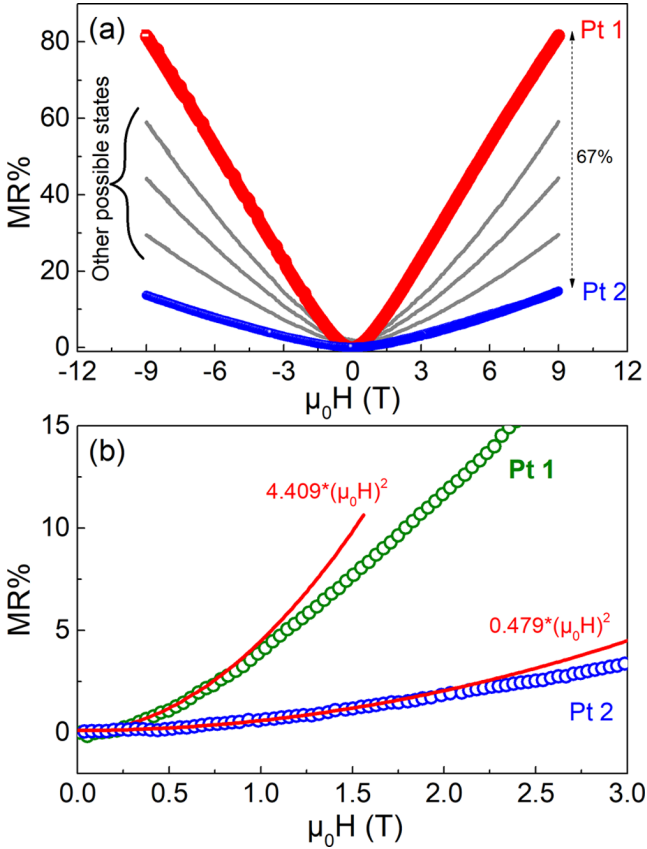

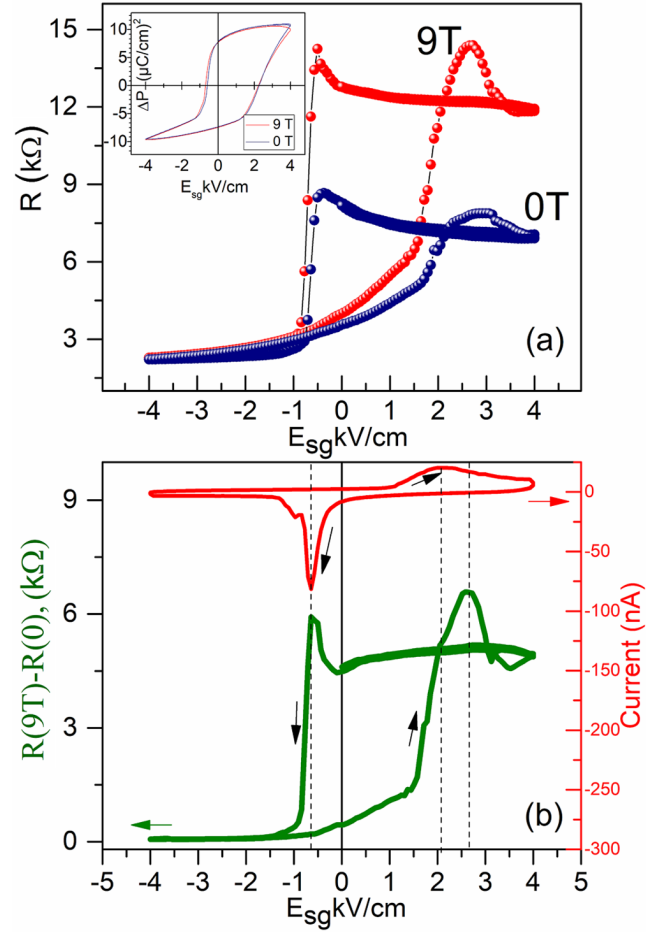

gives rise to higher carrier mobility and magnetoresistance.[3,33] Indeed, for point 1 (Fermi energy level near the Dirac point), the quadratically varied MR region shrinks to lower magnetic fields and enhances the linear MR dependence.

**2.3. Magnetic Control of Transconductance.** The variation of graphene resistance with gate voltage is shown for constant magnetic fields of 0 and 9 T in Figure. 4a. The

Figure 3. FE control of magnetoresistance. Magnetoresistance of graphene at the FE remanence (points 1 and 2) (a) along with other possible MR isotherms for subcoercive set of remanent FE states. MR effect fitted quadratically at low magnetic fields (b).

Figure 4. Magnetic field effect on graphene transconductance. The electroresistive (transconductance) loop depends on the magnetic field (a), while there is a negligible magnetic field effect on the FE loop (inset). Change in the magnetoresistance (b) (left scale) and its correlation with the source gate FE current (right scale).

For all the cases, MR shows positive values and varies quadratically at low magnetic field, and then it deviates to a linear dependence at larger magnetic fields. This behavior originates from the scattering of charge carriers which deviates their paths by Lorentz force due to an applied magnetic field (Drude model). In this case, MR = $\mu^2 H^2$ (where $\mu$ is the mobility of the charge carrier and $H$ is the applied magnetic field). The low magnetic field quadratic fitting in Figure 3b estimates the hole mobility at points 1 and 2 as 20371 and 6928 cm$^2$/(V s) respectively, which indicates the very large (over 13k cm$^2$/(V s)) difference in charge mobility between the FE remanent states. When Fermi level is near the Dirac point, it decreases the charge density in the graphene channel and both types of charge carriers participate in charge transport, and the presence of electron and hole puddles enormous magnetic field effect is observed with MR being higher for the up polarization state (P1) and lower for the down polarization state in accordance with MR loops. The direct magnetoelectric coupling of the substrate is negligible because FE loops are found not to be affected by magnetic field (Figure 4a (inset)). Notably the transconductance loop is magnified by 86%, which is ~5% more than expected from the MR loop measured at zero voltage at P1 (Figure 3a). This difference can be ascribed to the contribution of leakage current during electric field cycling in the latter case as a result of minor Ohmic conductivity. The charge-mobility-driven mechanism behind this observation can be confirmed by the correlation between the extracted magnetic-field-induced variation (difference between $H$ = 9 T and $H$ = 0 T loops) and source-gate FE current (Figure 4b).

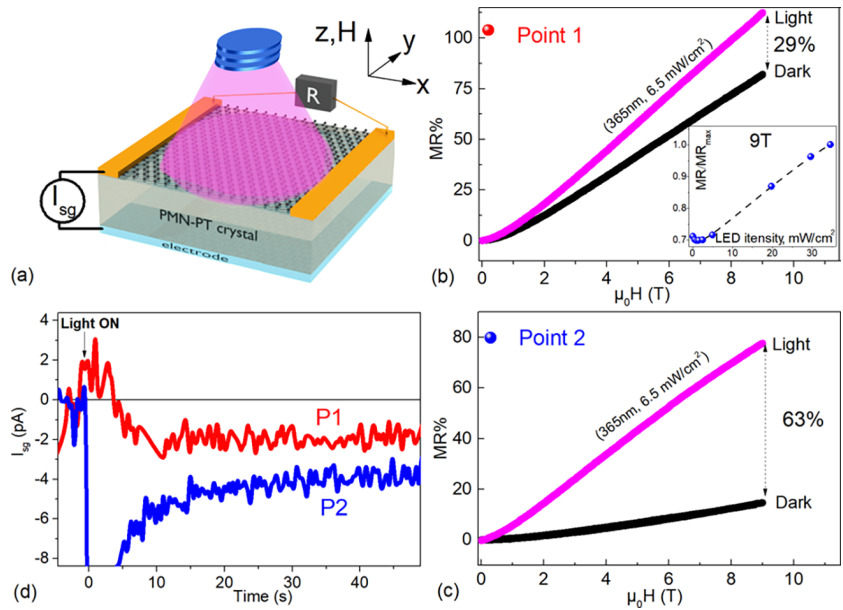

Figure 5. Bandgap light illumination effect on magnetoresistance. Illustration of experiment (a). Magnetoresistance of graphene at FE remanent points 1 (b) and 2 (c) in darkness and under continuous light. The inset to Figure 1b illustrates the relative magnetoresistance change as a function of 365 nm light intensity taken at 9 T. Time dependence of the PV current for different FE remanent states (d).

For a negative voltage sweep, the largest magnetoresistance is observed exactly at the FE dipole reorientation (maximum of charge mobility). For the positive electric fields, the magneto-resistive peak is shifted toward higher values, indicating space charge movement inside the sample. Indeed, the FE loop shows a larger leakage contribution with related vague saturation for the positive electric field values (Figure 2b).

**2.4. Optical Control of Magnetoresistance.** Following how an electric field dopes the graphene from the side of a ferroelectric, a similar function can be obtained in principle by other methods (for example, temperature or pressure[32]), which leads to FE polarization change and hence modification in graphene charge mobility. Because the PMN-PT substrate is known to be photovoltaic near the morphotropic phase boundary region,[34−36] we can use light with a bandgap energy to modify electrostatic doping and consequently the magneto-resistance of the graphene. In Figure 5 the latter was measured at two remanent states (point 1 and point 2) by applying continuous light irradiation ($\lambda$ = 365 nm, 6.5 mW/cm$^2$). In both remanent points 1 and 2 the magnetoresistance increases under light (Figure 5), reaching 100% for the remanent state of point 1. However, as can be seen from Figure 5c the light effect on MR is more than 2 times larger for the remanent point 2, reaching 63% modification. At this remanent point 2, obtained after negative poling, the graphene electrode attracts more holes (reducing charge mobility) to the graphene, leading to lower resistance and lower magnetoresistance in darkness. However, under light the PV current increases (Figure 5d), providing new free charge generation and accumulation on both sides of the ferroelectric driven by an internal electric field, which impacts the mobility due to the modulation of the Fermi energy level.

The observed light effect on the MR curves, therefore, is mostly dominated by the PV effect of the substrate. The latter is known to be asymmetric with respect to electric field poling as observed in many PV ferroelectrics[37−41] and can be explained by FE bias-induced asymmetric stresses.[32] Under such circumstances, the PV effect at point 1 must be lower than that in point 2, which is indeed experimentally observed (Figure 5d). In presence of the bandgap light, the larger PV effect at point 2 modulates the Fermi level more efficiently as compared to point 1, leading to a larger light effect on the variation of MR. However, for point P1, the Fermi level is closer to the Dirac point in the dark due to the "up" polarization of the ferroelectric. Therefore, by shining light, a further shift toward the Dirac point occurs, exhibiting higher MR effect magnitude surpassing 100%. Furthermore, the photovoltaic (PV) current, incorporating both pyroelectric and PV contributions,[42] may consequently exhibit a nonlinear dependence on light intensity.[43] This assertion is substantiated by the observed nonlinear magnetoresistance dependence in relation to light intensity, as illustrated in Figure 5b (inset).

**2.5. Optical Writing of Graphene Magnetoresistive Response.** Taking advantage of FE property of the substrate and the PV ability to change the polarization state by light,[19,44] one can write a permanent state to the graphene doping level by illumination and, therefore, write a permanent change to the MR response. Figure 6a demonstrates graphene resistance modification by UV light pulses as a function of time, where the initial resistance "A" and after illumination "B" levels are obtained, optically reproducing the synaptic memory.[45]

Therefore, the optical writing behavior of ferroelectric polarization as well as graphene resistance can be used to tune the wide range of MR variation in this device. Depending upon the intensity of light and pulse duration we can dope differently the graphene channel at different polarization states of the ferroelectric.[43] As a result, the magnetoresistance can be written optically. It has to be noted that sub-bandgap wavelengths can still affect the ferroelectric (FE) state through pyroelectric depolarization;[43] however, higher-energy light is more advantageous, as it involves free charge generation. As shown in Figure 6, the resistance of graphene is written to a higher value by using 2 optical pulses (17 mW/cm$^2$, during 100 ms). The light pulses change polarization and create photogenerated charges which are distributed along the polarization direction of [001], affecting the graphene resistance. We measured MR for point A, in the dark condition at the up-polarization state and at point B after pulsing light, and it is observed that the MR value is higher for point B compared to point A: i.e., MR is also written to higher values along with graphene resistance by an optical pulse.

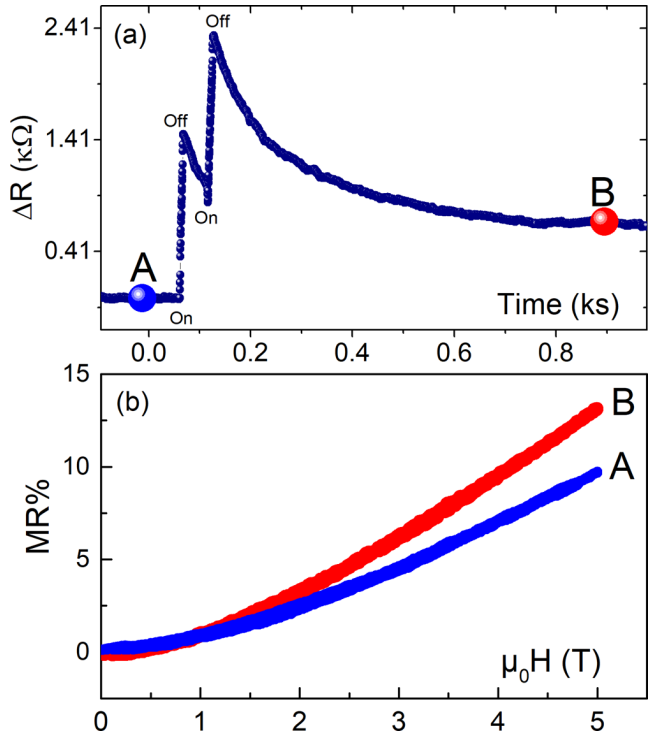

**Figure 6.** Optical writing of the MR response. Light can write the doping-determined graphene resistance level (a) with a larger MR effect with respect to the ground state (b).

## 3. CONCLUSIONS

In summary, first, an electrical method to modulate and write the MR property to graphene was demonstrated by deterministic tuning of the electrostatic doping level via the FE substrate. When optimized, this possibility can lead to on-demand MR response realization, where even different MR areas can be designed within the confines of a single device using FE domains. The underlying mechanism relies on electrostatic manipulation of free charge carrier density[43] and related mobility (i.e., larger mobility results in a larger MR effect). The MR magnitude is clearly more pronounced closer to the Dirac point, and our approach can tune an electron−hole plasma state near this region.[3] Moreover, in a similar way by exploiting the PV effect of the FE substrate one can affect the electrostatic equilibrium at the graphene/FE interface leading to even light-induced MR control. Because of the FE property, the exposure to light can result in a stable memory with synaptic functions[45] leading to an optical possibility of MR property writing and erasing.[19,43] Therefore, the on-demand and nonvolatile MR property control by either electrical or optical means or by the combination of the two should advance 2D-based devices for electro-optical and even all-optical applications.

## 4. METHODS

**4.1. Photoferroelectric Substrate.** The Pb[($Mg_{1/3}Nb_{2/3}$)$_{0.70}$-Ti$_{0.30}$]O$_3$ (PMN-PT30%) crystal with dimensions of 1.49 × 0.78 × 0.3 mm$^3$ was cut from plates of (001) orientation supplied by Crystal-GmbH (Germany). This ferroelectric compound belongs to the PMN-PT family of low-electric-field switchable crystals reported to manifest a large scale of interesting properties ranging from electro-optic[46−48] and piezoelectric[49−52] to photovoltaic[32,34,35,53] and photostrictive.[36,54]

**4.2. Graphene/Electrode Preparation.** Chemical vapor deposited (CVD) graphene was employed as a large-area highly sensitive surface charge sensor.[55] The service of CVD graphene monolayer deposition was ordered from Graphenea on the custom substrate (001) crystal plane with postannealing. The source and drain gold electrodes of 10 nm of Ti and 40 nm of Au were deposited at the edges of the graphene layer by E-beam evaporation and wired using silver conductive paste. To protect the graphene−FE interface, no wet steps were involved either in sample cutting or in electrode deposition to protect the interface. The in-plane source-drain resistance of graphene at a distance of 1.24 mm between the electrodes was measured as a function of gate voltage (Figure 2a).

**4.3. Ferroelectric Measurements and Photoexcitation.** The FE loop (Figure 2b) was measured along the [001] direction (Figure 2a) using a Keithley 6517B electrometer as a quasi-static loop tracer (at 0.01 Hz) performing current integration with respect to time, similar to the method described in ref 56. The obtained charge was divided by the electrode area to obtain polarization. The FE cycle was recorded in the absence of light by sweeping the electric field from zero to −4 kV/cm followed by a −4 to +4 kV/cm sweep and then back to zero with a step of 66.6 V/cm. The measurements have been performed at 300 K inside the PPMS cryostat at an average pressure of 20 mTorr. For optical control we have used a near-bandgap light excitation[57] provided by a Model M365FP1 365 nm light-emitting diode (LED) (Thorlabs) with 9 nm bandwidth calibrated with a Thorlabs power meter (Model PM100USB).

## AUTHOR INFORMATION

**Corresponding Author**

  B. Kundys − Université de Strasbourg, CNRS, Institut de Physique et Chimie des Matériaux de Strasbourg, UMR 7504, Strasbourg F-67000, France; orcid.org/0000-0001-6324-5302; Email: kundysATgmail.com

**Authors**

  Krishna Maity − Université de Strasbourg, CNRS, Institut de Physique et Chimie des Matériaux de Strasbourg, UMR 7504, Strasbourg F-67000, France

  Jean-François Dayen − Université de Strasbourg, CNRS, Institut de Physique et Chimie des Matériaux de Strasbourg, UMR 7504, Strasbourg F-67000, France

  Bernard Doudin − Université de Strasbourg, CNRS, Institut de Physique et Chimie des Matériaux de Strasbourg, UMR 7504, Strasbourg F-67000, France

  Roman Gumeniuk − Institut für Experimentelle Physik, TU Bergakademie Freiberg, Freiberg 09596, Germany; orcid.org/0000-0002-5003-620X

**Author Contributions**

B.K. initiated the idea of experiment. K.M. conducted measurements under the supervision of B.K. and J.-F.D. The funding was obtained by B.D., J.-F.D., and B.K. All authors contributed to the analysis of the results and scientific discussions. The manuscript was written by B.K and K.M. through contributions of all authors.

## ACKNOWLEDGMENTS

We acknowledge the financial support from IdEx 2022 - AAP Recherche Exploratoire program of Strasbourg University, ANR MixDFerro grant (ANR-21-CE09-0029), and chair program of Institut Universitaire de France. We also thank N. Beyer, and F. Chevrier for technical help, and J. Robert for useful discussions.